\documentclass[twocolumn,prl,showpacs,floatfix]{revtex4}%
\usepackage{graphicx}
\usepackage{bm}
\usepackage{dcolumn}
\usepackage{amsmath}
\usepackage{longtable}
\usepackage{times}
\usepackage{amsfonts}
\usepackage{amssymb}%

\begin{document}

\title{Interplay between localized and itinerant $d$ electrons in a frustrated
metallic antiferromagnet, $2H$-AgNiO$_{2}$}
\author{A. I. Coldea$^1$, A. Carrington$^1$, R.
Coldea$^{1}$,  L. Malone$^1$,   A.F. Bangura$^1$, M. D. Johannes$^2$, I. I. Mazin$^2$,\\
 E.A. Yelland$^1$, J. G. Analytis$^1$, J.A.A.J. Perenboom$^4$, C. Jaudet$^5$,
 D. Vignolles$^5$, T. Sorgel$^3$, M. Jansen$^3$}
\affiliation{$^{1}$H. H. Wills Physics Laboratory, Bristol
University, Tyndall Avenue, BS8 1TL, United Kingdom}
\affiliation{$^{2}$Code 6393, Naval Research Laboratory,
Washington, D.C. 20375, USA}
\affiliation{$^{3}$Max-Planck-Institut fur Festkorperforschung,
Heisenbergstr. 1, 70569 Stuttgart, Germany}
\affiliation{$^{4}$High Field Magnet Laboratory, IMM, Radboud
University, 6525 ED Nijmegen, The Netherlands}
\affiliation{$^{5}$Laboratoire National des Champs Magnetiques
Pulses, 31400 Toulouse, France}

\begin{abstract}
We report the electronic and magnetic behaviour of the frustrated
triangular metallic antiferromagnet $2H$-AgNiO$_{2}$ in high
magnetic fields (54~T) using thermodynamic and transport
measurements. Here localized $d$ electrons are arranged on an
antiferromagnetic triangular lattice nested inside a honeycomb
lattice with itinerant $d$ electrons. When the magnetic field is
along the easy axis we observe a cascade of field-induced
transitions, attributed to the competition between easy-axis
anisotropy, geometrical frustration and coupling of the localized
and itinerant system. The quantum oscillations data suggest that
the Fermi surface is reconstructed by the magnetic order but in
high fields magnetic breakdown orbits are possible. The itinerant
electrons are extremely sensitive to scattering by spin
fluctuations and a significant mass enhancement ($\sim3$) is
found.

\end{abstract}

\date{\today}
\maketitle

Strongly-interacting electrons in triangular layers display a
variety of correlated phases stabilized by the frustrated lattice
geometry; for example, spin liquid in a triangular organic Mott
insulator in the close proximity of pressure-driven superconducting
state \cite{Kanoda2003} or superconductivity in water-intercalated
Na$_{x}$CoO$_{2}$ \cite{Takada2003}. High magnetic fields can be
used as a tuning parameter to manipulate the electronic ground state
and drive transitions to novel phases and here we explore such
physics in the metallic triangular antiferromagnet,
$2H$-AgNiO$_{2}$, proposed to realize a novel paradigm for charge
order in orbitally-degenerate weakly-itinerant
systems~\cite{Wawrzynska2007,Mazin2007}.
Interestingly, below the charge order transition at 365~K the
system remains metallic, and a rather unusual electronic ground
state is proposed by band-structure calculations: a triangular
lattice of electron-rich and localized Ni1 sites (Ni$^{2+}$,
$S=1$) nested inside a honeycomb lattice of Ni$^{3.5+}$ where
electrons remain itinerant (see Fig.1f). A novel magnetic ground
state is observed at low temperatures where the spins of the
localized Ni1 sites order in a collinear structure of alternating
stripes, whereas the itinerant electrons remain magnetically
unordered \cite{Wawrzynska2007}. In AgNiO$_{2}$ the spin
anisotropy and magnetic interactions are sufficiently low that a
large part of the whole magnetic phase diagram in field is
experimentally accessible and here we reveal a cascade of
field-induced phases, with itinerant electrons being strongly
affected by the spin fluctuations near the phase boundaries, with
reconstructions of the Fermi surface and magnetic breakdown orbits
becoming possible at high field.

\begin{figure*}[ptbh]
\centering
\includegraphics[width=5.7cm,height=8.9cm]{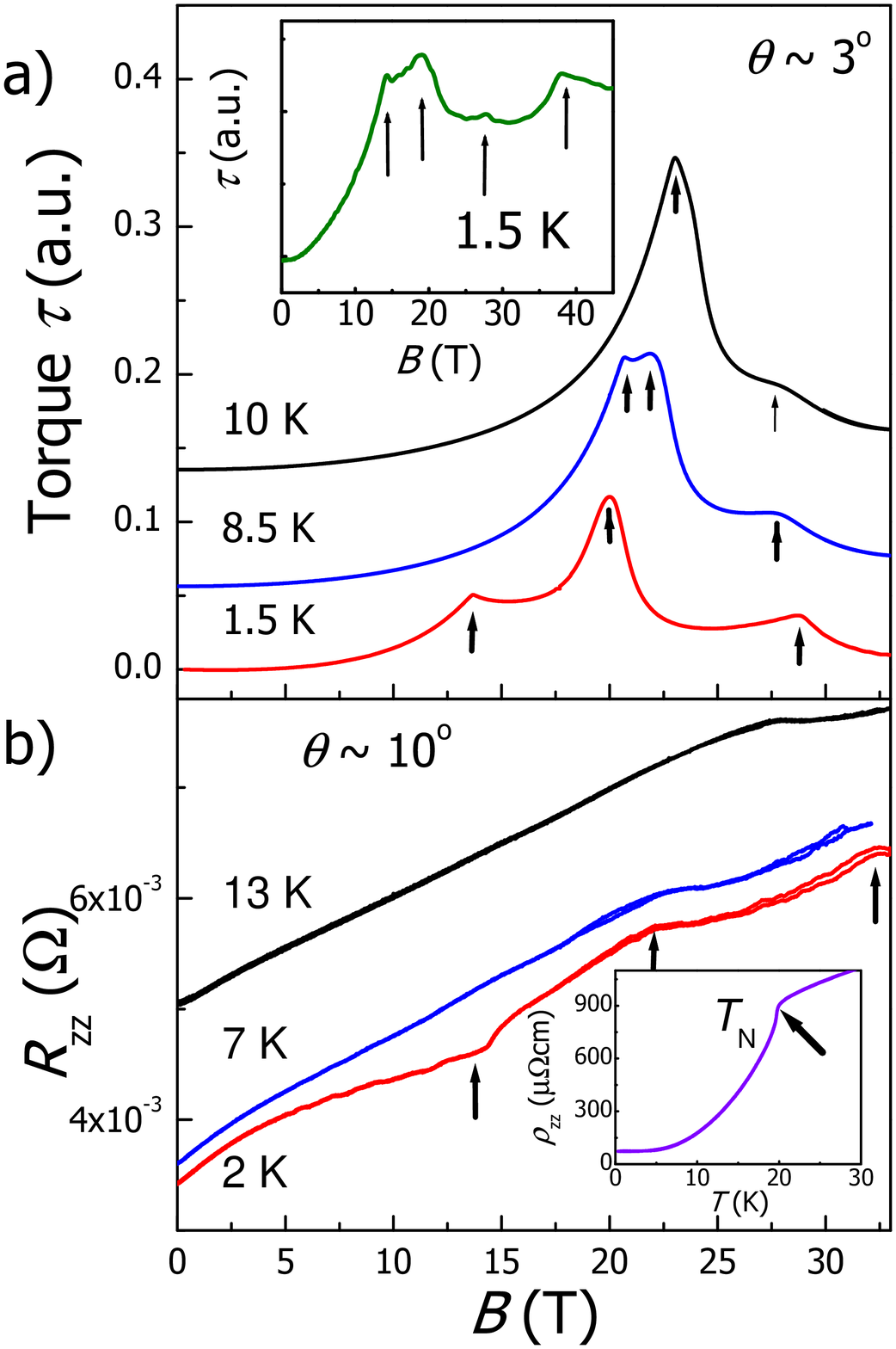}
\includegraphics[width=5.7cm,height=8.9cm]{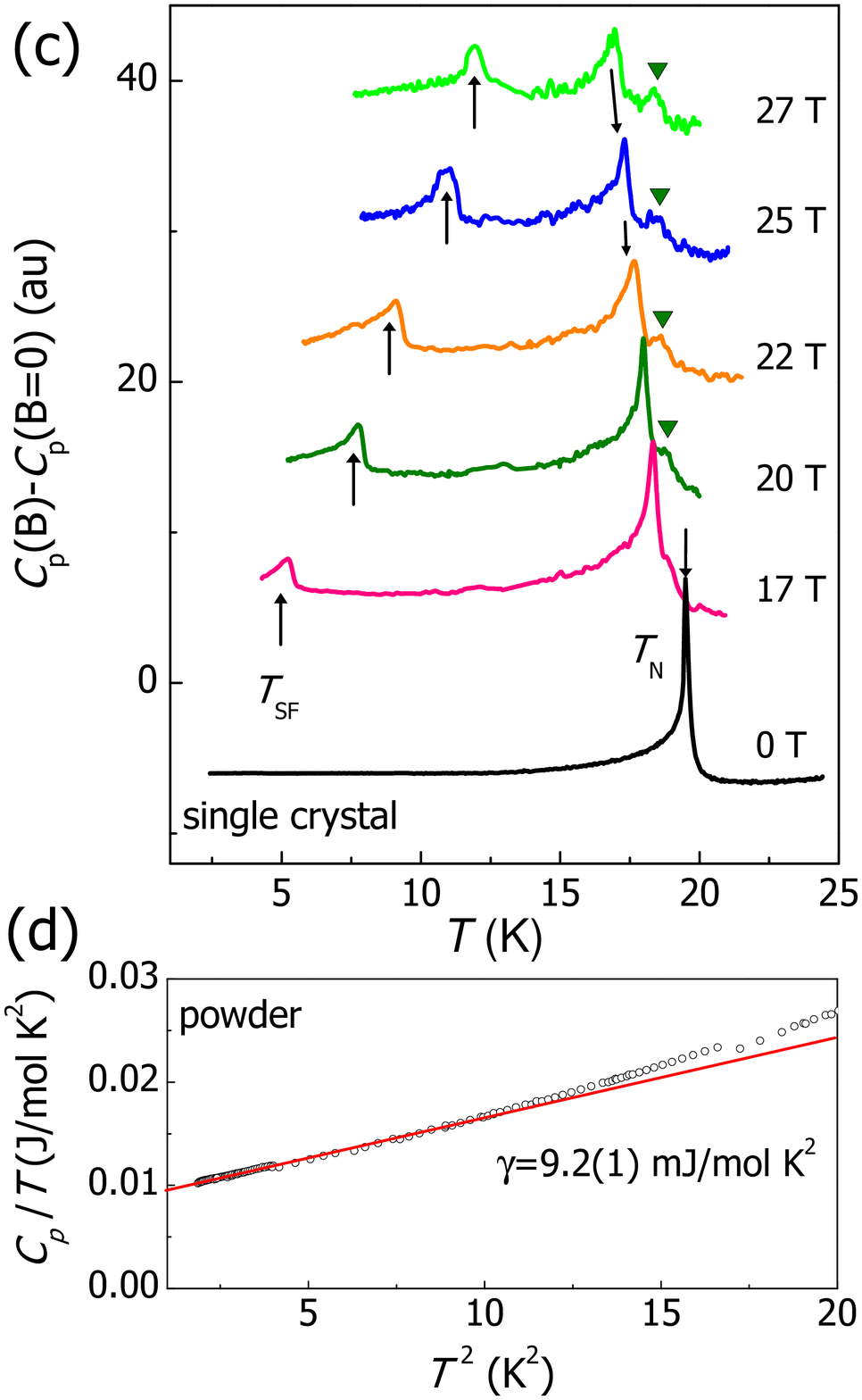}
\includegraphics[width=5.7cm,height=9.3cm]{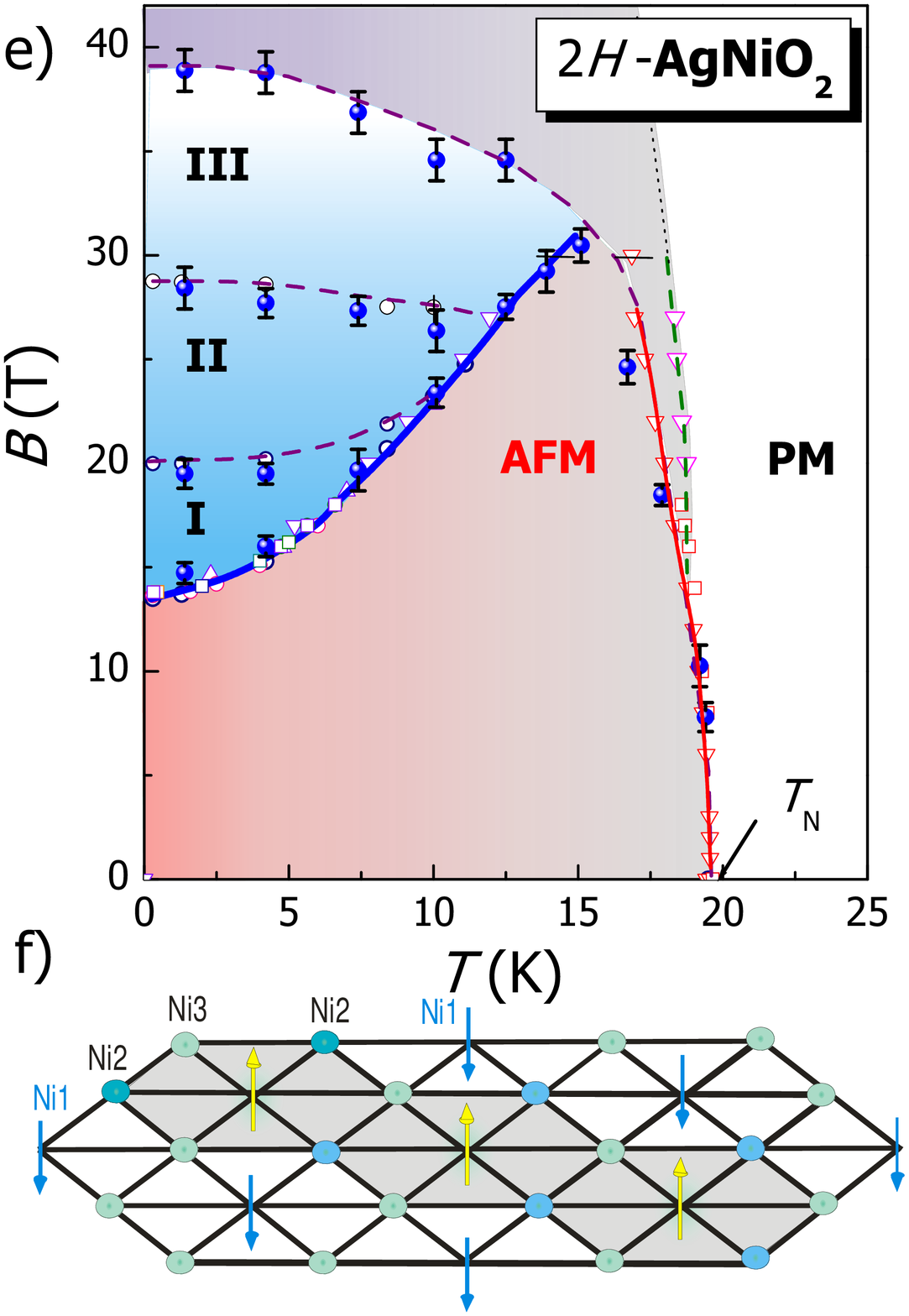}\caption{(colour
online) Field dependence of (a) the magnetic torque response, and (b) the
interplane resistance at fixed temperatures when the magnetic field is close
to the $c$ axis ($\theta\approx3^{\circ}$ for torque measurements and
$\theta\approx10^{\circ}$ for transport measurements)(sample B). The top inset
show torque data in pulsed magnetic fields on a sample D and bottom inset
shows the zero-field interplane resistance at low temperatures. Arrows
indicate anomalies attributed to different magnetic transitions. (c) The
relative variation of specific heat in different constant magnetic fields for
sample C and (d) the low temperature extrapolation of $C_{p}/T$ to in zero
field for a powdered sample. The arrows indicate various transitions and the
solid triangles corresponds to an additional transition in high fields. (e)
Proposed phase diagram of $2H$-AgNiO$_{2}$ from torque magnetometry (circles),
specific heat (triangles), transport (square). The solid and dashed lines
indicate boundaries between different magnetic phases: antiferromagnetic
(AFM), high-field stabilized phases I, II, III, and paramagnetic (PM). A
possible high field and high-temperature transition is also indicated (open
triangles). The collinear magnetic structure (AFM phase) of Ni1 spins (arrows
pointing up and down along $c$-axis) and honeycomb network for itinerant
electrons is shown in (f).}%
\label{raw_data}%
\end{figure*}

Torque measurements are well suited for investigating both changes
in the magnetic ground state, manifested in kink anomalies in the
torque response, as well as the Fermi surface (FS) of the itinerant
electrons, via observation of quantum oscillations in the
magnetization (de Haas-van Alphen effect). Torque, ${\bm{\tau}}={\bm
m} \times{\bm B}$, in magnetic materials is caused by anisotropy,
and measures the misalignment of the magnetization, ${\bm m}$, with
respect to the applied magnetic field ${\bm B}$. We measured the
overall torque response using a sensitive piezo-resistive cantilever
technique. Specific heat was measured using a purpose built
calorimeter which uses both a long relaxation and an \emph{ac}
relaxation method. Hexagonal-like single crystals (typical size
$\sim120 \times100 \times10~\mu$m$^{3}$, 6 crystals were
investigated) were grown using a solid-state route under high oxygen
pressures \cite{growth}). The residual interlayer resistivity ratio
was $\sim250$. Measurements were performed at low temperatures
(0.3~K) on different crystals in steady fields up to 18~T in
Bristol, 33~T at the HFML in Nijmegen and in pulsed fields up to
56~T at the LNCMP,
Toulouse.

Figs.~\ref{raw_data}a-b) show the field and temperature dependence of the
torque signal and interlayer resistivity, respectively, performed with the
magnetic field aligned close to the high symmetry easy $c$ axis ($\theta=0$
when $B || c$). At low temperatures the torque signal varies as $H^{2}$ in low
fields suggesting that magnetization increases linearly with magnetic field.
In higher fields we observed a series of kinks in torque (13.5~T, 20~T, 28.8~T
and 38~T) which we attribute to magnetic phase transitions, similar anomalies
have been seen at field-induced transitions in other uniaxial antiferromagnets
\cite{Kawamoto2008}. The locations of the torque anomalies correlate closely
with kinks seen in the interplane conduction, see Fig.~\ref{raw_data}b)
(slightly shifted in field due to a difference in sample orientation),
suggesting that the itinerant $d$ electrons are highly sensitive to the
changes in the magnetic order pattern.

A phase diagram constructed on the basis of torque, resistivity and specific
heat measurements on several crystals is shown in Fig.~\ref{raw_data}e). The
specific heat data shown in Fig.~\ref{raw_data}c confirm the transition
between the paramagnetic (PM) and AFM phase at the N\'{e}el temperature
$T_{\mathrm{N}}$=19.5~K and a further strong anomaly at a lower temperatures
(likely first-order), which coincides with crossing the boundary separating
the AFM and the higher field phase (I). In the limit of our experimental
resolution (due to extremely small samples) we cannot detect clear signatures
for the other two transitions (between phases I to II, and II to III), clearly
observed both in the torque and transport measurements (Fig.~\ref{raw_data}%
a-b); this could be due to the fact that the specific-heat scans
were performed mainly at constant temperature and transition
boundaries are nearly flat and/or that these transitions are
related to magnetic effects that do not involve a significant
change in entropy.

\begin{figure}[ptbh]
\centering
\includegraphics[width=7cm]{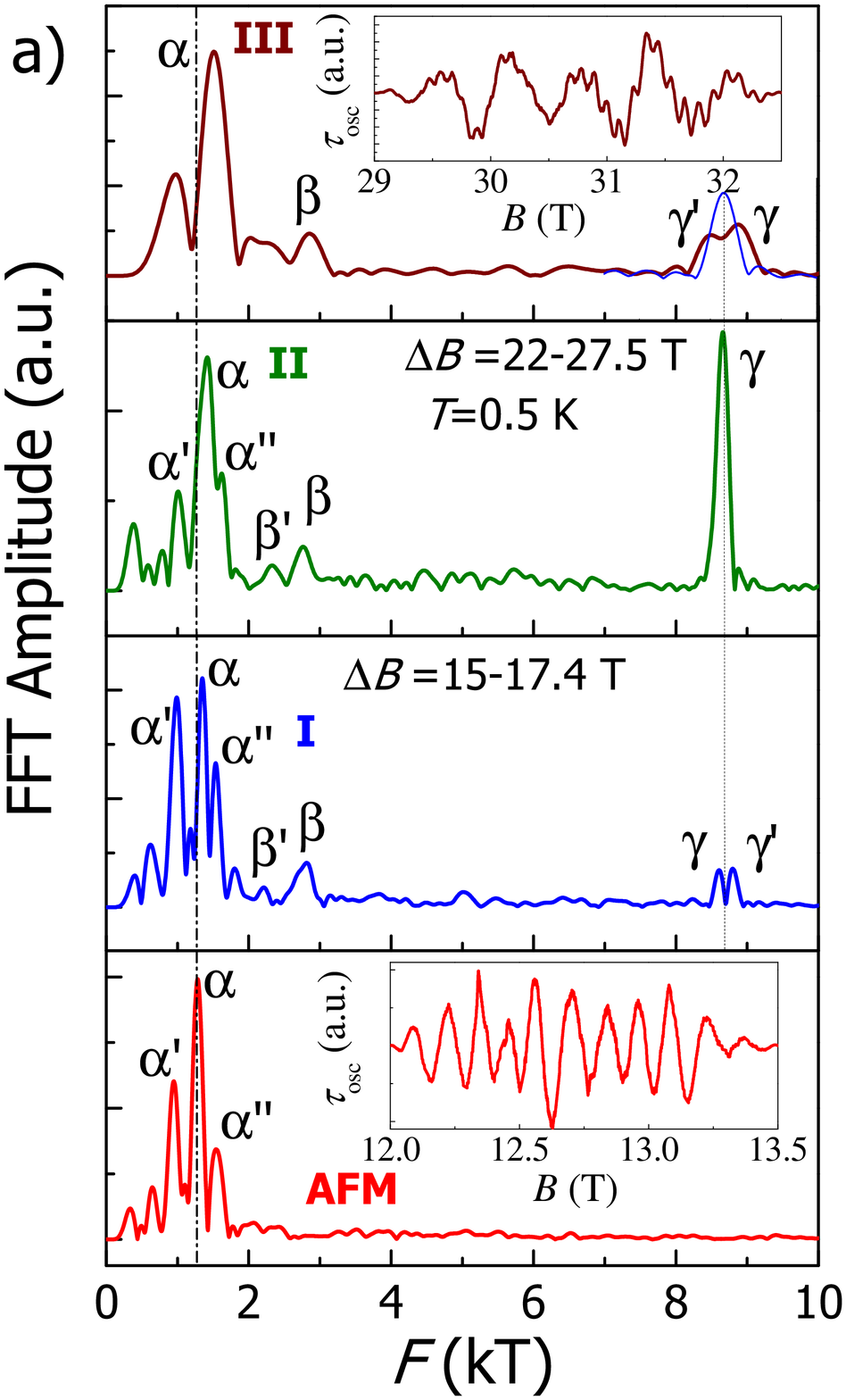}
\includegraphics[width=7cm]{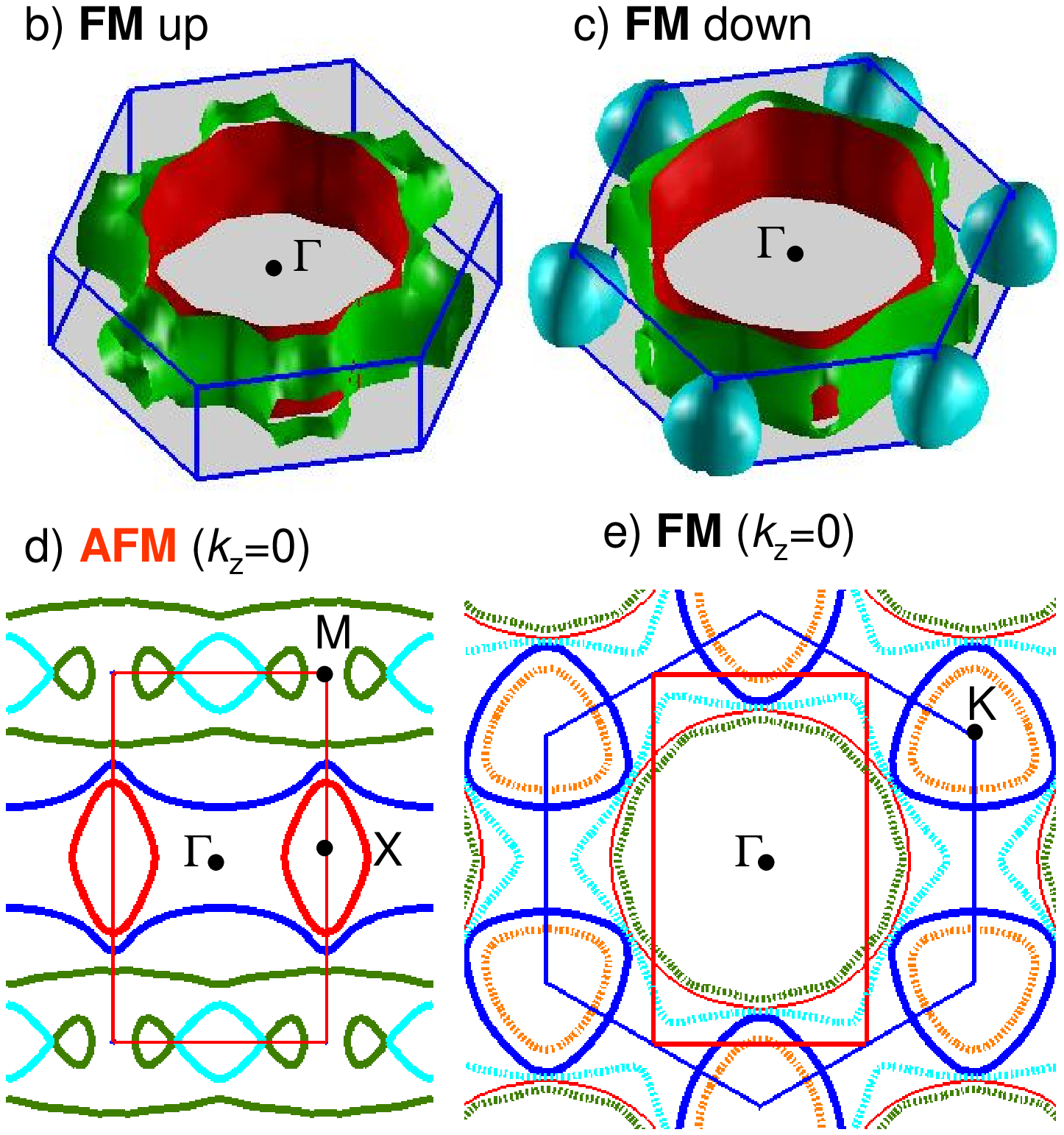}\caption{(color online) (a)
Fourier transformed spectra for different magnetic phases observed when the
magnetic field is aligned close to the $c$ axis ($\theta\approx3^{\circ}$ and
$T=0.5(1)$~K). The raw dHvA data are shown as insets. Similar FFT windows
(1/$\Delta B$) were used up to 28~T (top panel shows a spectra of phase III
compared with phase I for a similar field range (thin line)). The vertical
lines are guides to the eye. (b) and (c) FS of the fully polarized
configuration (see Ref.\onlinecite{Wawrzynska2007}) and slices of the (d) AFM
and FM Fermi surface at the $\Gamma$ point when the magnetic field is pointing
along the $c$ axis. The predicted orbits are shown (solid and dotted lines)
inside the AFM (solid rectangle) and non-magnetic (hexagonal) Brillouin zone.}%
\label{FFTs}%
\end{figure}

The richness of the observed phase diagram is a manifestation of
the complexity of magnetic interactions in the system. Besides the
frustrated nature of triangular planar magnets, AgNiO$_{2}$ is
unique in the sense that it has several different magnetic
interactions of disparate nature, but of the same scale. Indeed,
since Ni1 ions are rather far apart, the superexchange between
them is weak and cannot entirely dominate the physics of the
system. On the other hand, the experimentally measured magnetic
anisotropy gap is surprisingly large ($\sim1.7$ meV
\cite{Elisa2008}) for a closed-shell ion, emphasizing the
significant role of hybridization with Ni2,3 for magnetic
interactions. Since Ni1 ions are embedded in a metal (Ni2 and
Ni3), they are subject to RKKY interaction, which, unlike
superexchange, decays slowly with distance. For example, RKKY
may provide non-negligible $nnn$ Ni1 exchange, even to those ions
that are too far apart for superexchange. Last but not least, the
calculations show a small, but finite (0.1-0.2 $\mu_{B})$ moment on
the itinerant and inherently nonmagnetic Ni3 sites. The Hund rule
coupling on these sites provides an additional incentive for Ni1s
to order in such a way that the Ni3 moment be nonzero by symmetry
(for instance, favoring the observed structure over the $nn$-only
Heisenberg 120$^{o}$ structure). The scale of this interaction is
set by the Hund's rule coupling energy on Ni3, $Im^{2}/4,$ where
the Stoner factor $I\sim0.6-0.8$ eV and $m\sim0.1-0.2$ $\mu_{B},$
and appears to be a few meV.  By a similar mechanism, the Hund's
energy of induced moments generates ferromagnetism in SrRuO$_{3}$
\cite{Mazin1997}.
Note that a coupling between the conduction (Ni2,3) electrons and
the local moments is evidenced by the significant drop in
resistivity below $T_{\mathrm{N}}$, as the result of suppression
of electron scattering by low-energy spin fluctuations when a spin
gap opens (see inset Fig.~1b). Moreover, the magnetoresistance
shows changes in slope in the vicinity of the magnetic transitions
(Fig.~1e), indicating significant scattering by the spin
fluctuations close to these transitions. Finally, interplanar
coupling (which also proceeds $via$ the weakly-magnetic Ni3 ions
\cite{Wawrzynska2007}), while small, is not negligible either
compared to other, in-plane interactions.

Even an oversimplified model of the localized Ni1 sublattice that
includes only the on-site anisotropy, the $nn$ exchange $J$, and a
small $nnn$ exchange $J'$,  already demonstrates a surprising
richness. Aside from the observed zero-field AFM collinear phase
(Fig.1f) it allows for multiple phases in field, such as a
modified version of the AFM phase, AFM$'$,  where the rows of up
spins remain ferromagnetically aligned along the field
$\uparrow\uparrow\uparrow\uparrow$ and the rows of down spins
develop an additional weak transverse antiferromagnetic order as
they tilt away from the easy axis
$\swarrow\searrow\swarrow\searrow$. The AFM$'$ configuration may
be a candidate for phase I in Fig.~1e \cite{Seabra2009}. The
mean-field critical field for the transition between AFM and
AFM$'$
is $B_{c}=\Delta/g\mu_{B}$; note that the experimentally observed
$B_{c}=13.5$~T (at low temperatures) agrees with this estimate,
using $\Delta=1.7$~meV from neutron data \cite{Elisa2008} and
assuming $g=2.17$. One can show that at the mean-field level
\cite{Seabra2009} the AFM$'$ phase exists in a range of fields
above $B_c$ for $J'>\Delta/12S$, and is replaced by a honeycomb
arrangement (spins aligned with the field on the honeycomb lattice
with an anti-aligned spin on the central site). Other more complex
phases manifest at higher fields, so that even in this highly
simplified model there are multiple phases that can be stabilized
by an external field, as seen in Fig.~1e.
However, a more realistic model for AgNiO$_2$ would also need to
account for the polarization of the itinerant electrons which brings
in a Hund's rule energy term that can also distinguish between
competitive phases. It is the itinerant Ni3 sites that provide the
exchange link between successive layers so changes in the Ni3
polarization in field could also affect the 3D stacking of the
magnetic order of the main Ni1 moments \cite{Wawrzynska2007}, and
this might explain some of the higher-field transitions in
Fig.\ref{raw_data}e.

If the itinerant $d$ electrons are indeed important for the magnetic
interaction and sensitive to the external field, one expects an interesting
evolution of the Fermi surfaces with the field. We monitored this evolution
using quantum oscillations in magnetic field. The total torque signal is
dominated by the localized moment magnetism away from the easy axis and
various magnetic phases have strong angular dependence in field
\cite{ColdeaA2008}, so here we concentrate on the behavior for fields close to
the easy axis. De Haas-van Alphen oscillations are obtained by subtracting a
fifth order polynomial from the total torque in different magnetic regions
(see insets of Fig.2a). The Fourier transform allows us to identify the
corresponding frequencies related to the extremal areas of the Fermi surface
by the Onsager relation, $F=(\hbar/2\pi e)A_{k}$.

In the AFM phase the FFT spectra consist of closely packed frequencies below
2~kT with the most intense peak at $\alpha\approx1.3$~kT, see Fig.~2a). Above
13.5~T (phase I) additional split high frequencies appear, $\gamma\approx
8.6$~kT, as well as $\beta\approx2.8$~kT (and a weak $\beta^{\prime}$). At
much higher fields the spectra is composed of a similar number of frequencies,
suggesting that the Fermi surface evolves rather smoothly above the first
transition at 13.5~T. The $\alpha$ pocket, occupies an area of $\sim8$\%
whereas $\gamma$ occupies $\sim46$\% of the non-magnetic Brillouin zone
(hexagonal contour in Fig.~2d). In the AFM phase the Fermi surface is expected
to be reconstructed by the collinear magnetic order ($A_{AFM}$=$0.88
\times10^{20}$ m$^{-2}$) and the $\gamma$ pocket, if present, would occupy
almost this entire magnetic zone area; however the $\gamma$ pocket is not
observed in the AFM phase suggesting that the Fermi surface is composed of
small pockets that may result from reconstruction in the magnetic Brillouin
zone \cite{breakdown}.

Fig.~2d shows the calculated Fermi surface in the AFM phase (using the same
computational methods as in Ref.~\onlinecite{Wawrzynska2007}). The FS is
formed of a large number of small pockets with the largest predicted extremal
areas corresponding to 1.7~kT (around the X point) which is in the range of
values observed experimentally for the $\alpha$ pocket. As the magnetic field
is tilted away from the easy axis (towards the $a$ axis) these low frequencies
show sizable angular dependence in agreement with their complex quasi-2D and
3D origin \cite{ColdeaA2008}.

In the high field phases, we consider the Fermi surface for a
spin-polarized ferromagnetic (FM) ground state, where all magnetic
spins are aligned (Figs.2b-d). This is also an approximation for
an \textit{unreconstructed} Fermi surface specific to AgNiO$_{2}$
(note that because Ni1 sites are localized and strongly-magnetic
means that a non-magnetic paramagnetic-like FS is not relevant
here). This spin-polarized \emph{large} Fermi surface is formed of
quasi two-dimensional warped cylinders with and without neck
orbits (centered around the $\Gamma$ point) with large areas that
can be assigned to $\gamma$ pockets. Around the K point there is a
quasi-three dimensional electronic pocket (of about 1~kT) similar
to the $\alpha$ pocket as well as orbits originating from the
necks of the large cylinders that could explain the presence of
$\beta$ pockets (see Fig.2e). Thus, qualitatively, in high
magnetic fields the observed FS pockets can be attributed to the
large and small bands of the FM calculations. However, we would
expect the new magnetic zone boundaries determined by the new
magnetic order in phases I, II and III to open up new gaps on the
Fermi surface. Observation of the high frequency $\gamma$ in all
these phases suggests that these gaps are small enough for the
electrons to tunnel through, as the probability of tunnelling
increases strongly in field, and the $\gamma$ frequencies would
correspond to magnetic breakdown orbits.

In a magnetic field and/or in the presence of the localized $d$
electrons ($\sim1.5\mu_{B}$) the electronic bands will be shifted
and/or split. As the system goes through different magnetic phases
we observed that the $\alpha$ pocket is slightly shifted towards
higher values (see Fig.2a) whereas the splitting of the large
$\gamma$ pockets varies. If this split results from the
spin-polarization of the quasi 2D bands (with similar shape
centered at the $\Gamma$ point in Fig.2e) by the local moments,
then the direct exchange coupling estimated as $\approx\hbar e
\delta F/(2 m^{*})$ \cite{Goodrich2006} varies from 1.62~meV in
phase I to 3.21~meV in phase III, similar in magnitude to
antiferromagnetic NdB$_{6}$ where the localized $f$ electrons are
coupled to the conduction electrons \cite{Goodrich2006}. This
relatively small direct exchange splitting could explain the
presence in the calculations of a small magnetic moment $\sim$0.1
$\mu_{B}$ on the itinerant Ni3 site \cite{Wawrzynska2007}.

In order to test whether electronic correlations are important we compare the
measured $\gamma$ of $9.2(1)$~mJ/mol $K^{2}$ extracted from specific heat data
(Fig.1d) with the calculated value of $3.36(1)$~mJ/mol K$^{2}$ for the AFM
phase. This gives a mass enhancement of $\sim2.6$, close to that found in
Ag$_{2}$NiO$_{2}$ \cite{Johannes2007,Ag2NiO2} or Sr$_{3}$Ru$_{2}$O$_{7}$
\cite{Borzi2004}, suggestive of strong spin fluctuations. Using the
Lifshitz-Kosevich formula \cite{Shoenberg1984} the effective mass for the
$\alpha$ pocket varies from $\sim2.3$~$m_{e}$ in the AFM phase to $\sim
3.2$~$m_{e}$ in phase I. The effective mass of the $\gamma$ pocket increases
only slightly with the magnetic field from $\sim6.0 ~ m_{e}$ (phase I) to
$\sim6.6 ~ m_{e}$ (phase III) reflecting the increase in magnetic
susceptibility with magnetic field.

In conclusion, we report new rich physics induced by the magnetic
field in $2H$-AgNiO$_{2}$ as a result of the competition between
strong easy-axis anisotropy, frustrated antiferromagnetic
interactions and coupling between localized and itinerant electrons.
The Fermi surface is reconstructed by the magnetic zones but in
higher fields the gaps opened at the zone boundaries are small
enough that the electrons can tunnel though. The corresponding
effective masses are enhanced by a factor of $\sim3$ due to strong
spin fluctuations. Similar physics could be relevant to other
systems, such as the parent compounds of superconducting iron
pnictides, where the rich physics observed can be determined by the
interplay between local moment magnetism and itinerant electrons
\cite{Zhao2009}.

This work is supported by EPSRC (UK) and AIC is grateful to the
Royal Society (UK) for financial support. Access to high magnetic
field facilities has been supported by EuroMagNET II under EU
contract.

\end{document}